\documentclass[12pt]{article}

\usepackage{amssymb}

\def\sech{\mathop{\textrm{sech}}\nolimits}
\def\sinh{\mathop{\textrm{sinh}}\nolimits}
\def\tanh{\mathop{\textrm{tanh}}\nolimits}

\begin{document}

\begin{center}
{\Large\bf Melnikov analysis for multi-symplectic PDEs}\\ \ \\
K.B. Blyuss\\
{\it Department of Mathematics and Statistics\\
University of Surrey, GU2 5XH, Guildford, UK}\\
E-mail: k.blyuss@surrey.ac.uk

\end{center}

\begin{abstract}
In this work the Melnikov method for perturbed Hamiltonian wave
equations is considered in order to determine possible chaotic behaviour in the
systems. The backbone of the analysis is the multi-symplectic formulation of the unperturbed PDE and its further reduction to
travelling waves. In the multi-symplectic approach two separate symplectic operators are introduced for the spatial and temporal variables, which
allow one to generalise the usual symplectic structure. The systems
under consideration include perturbations of generalised KdV equation,
nonlinear wave equation, Boussinesq equation. These equations are equivariant with respect to abelian subgroups of Euclidean group. It is assumed that the external
perturbation preserves this symmetry. Travelling wave reduction for the above-mentioned systems results in a four-dimensional system of ODEs,
which is considered for Melnikov type chaos. As a preliminary for the
calculation of a Melnikov function, we prove the persistence of a fixed point for the perturbed Poincare map by using Lyapunov-Schmidt reduction.
The framework sketched will be applied to the analysis of possible chaotic behaviour of travelling wave solutions for the above-mentioned PDEs
within the multi-symplectic approach.
\end{abstract}

\section{Introduction}

Recently it was shown how many nonlinear PDEs can be formulated in a
multi-symplectic form
\cite{3,4,5}.
This formulation assigns distinct symplectic structures to the
spatial and the temporal coordinates, thereby generalising the usual
Hamiltonian formulation. By means of the multi-symplectic approach
questions like stability of solitary waves, existence of generalised
basic state at infinity, equivariant properties of the solutions etc. can be
considered in a more general setting, yielding new results on these issues.

In order to analyse chaotic behaviour of travelling wave solutions to
Hamiltonian PDEs, Melnikov's method can be used. The problem with its
direct application is due to the symmetry in a multi-symplectic
formulation of these PDEs, which results in the presence of unit
eigenvalue among the spectrum of the Poincar\'e map. This complication
is solved by means of Lyapunov-Schmidt reduction.

Finally, we illustrate the application of the Melnikov method
to the study of chaotic behaviour in a perturbed Korteweg-de Vries
equation.

\section{General setting}

We start to consider a multi-symplectic PDE of the form \cite{5}:
\begin{equation}\label{1}
{\bf M}Z_{t}+{\bf K}Z_{x}=\nabla S(Z)+\epsilon S_{1}(Z,x-ct),\mbox{ }Z=\left(
\begin{array}{c}
U\\
V\\
W\\
\Phi
\end{array}
\right)\in\mathbb{R}^4,
\mbox{ }x\in\mathbb{R}
\end{equation}
where ${\bf M}$ and ${\bf K}$ are constant skew-symmetric matrices on
$\mathbb{R}^{4}$ and $S:\mathbb{R}^{4}\rightarrow\mathbb{R}$ is
sufficiently smooth ( at least twice continuously
differentiable). The perturbation $S_{1}$ is assumed to be a periodic
function of its last argument: $S_{1}(\cdot,x)=S_{1}(\cdot,x+T)$ and
also $C^{r}$, $r\geq 2$.

We suppose that the system (\ref{1}) is equivariant
with respect to a one-parameter Lie group, whose algebra is spanned by
the generator $\xi$. For the unperturbed case $(\epsilon=0)$,
multi-symplectic Noether theory provides the existence of the two
functionals $P(Z)$ and $Q(Z)$ such that \cite{3}
\begin{equation}
{\bf M}\xi(Z)=\nabla P(Z),\mbox{ }{\bf K}\xi(Z)=\nabla Q(Z),
\end{equation}
The state at infinity should satisfy \cite{4,5}:
\begin{equation}
\nabla S(Z_{0})=a\nabla P(Z_{0})+b\nabla Q(Z_{0})
\end{equation}
with $P(Z_{0})=\mathcal{P}$ and $Q(Z_{0})=\mathcal{Q}$ specified real
parameters, $a,b\in\mathbb{R}$.

A shape of an unperturbed solitary wave travelling at speed $c$,
$Z(x,t)=Z(x-ct)$, which is biasymptotic to this state should satisfy the equation
\begin{equation}
Z_{x}={\bf J}_{c}^{-1}\nabla H_{0}(Z),
\end{equation}
where $H_{0}(Z)=S(Z)-aP(Z)-bQ(Z)$, and ${\bf J}_{c}={\bf K}-c{\bf M}$ \cite{4,5}.

To study the existence of travelling waves and their chaotic
behaviour, we consider the dynamical system ( similar consideration
can be found in \cite{7})
\begin{equation}\label{5}
\frac{d}{dx}Z=f_{0}(Z)+\epsilon f_{1}(Z, x),\mbox{
  }Z\in\mathbb{R}^{4},\mbox{ }0<\epsilon\ll 1,\mbox{ }0<x<\infty,
\end{equation}
where $f_{0}(Z)={\bf J}_{c}^{-1}\nabla H_{0}(Z)$, and $f_{1}(Z,x)={\bf
  J}_{c}^{-1}S_{1}(Z,x)$.

The following hypotheses are imposed on the system:
\begin{description}
  \item[(H1)]
  \begin{description}
   \item[a)] {\it $f_{0}:\mathbb{R}^{4}\rightarrow\mathbb{R}^{4}$ is
       $C^{r}$ $(r\geq 2)$};
   \item[b)] {\it $f_{1}:\mathbb{R}^{4}\times
   S^{1}\rightarrow\mathbb{R}^{4}$ is $C^{r}$ $(r\geq 2)$}.
  \end{description}
\end{description}
The system (\ref{5}) can be rewritten as a following suspended
system:
\begin{equation}\label{6}
\left\{
\begin{array}{l}
\displaystyle{\frac{dZ}{dx}=f_{0}(Z)+\epsilon f_{1}(Z,\theta)},\\ \\
\displaystyle{\frac{d\theta}{dx}=\omega},
\end{array}
\right.
\end{equation}
with the frequency $\omega=2\pi/T$. Its flow
$\Phi_{t}^{\epsilon}:\mathbb{R}^{4}\times
S^{1}\rightarrow\mathbb{R}^{4}\times S^{1}$ is defined for all
$t\in\mathbb{R}$.
\begin{description}
 \item[(H2)]
 \begin{description}
  \item[a)] {\it The unperturbed system
\begin{equation}\label{7}
\frac{d}{dx}Z=f_{0}(Z)={\bf J}_{c}^{-1}\nabla H_{0}(Z)
\end{equation}
is Hamiltonian with energy $H_{0}:\mathbb{R}^{4}\rightarrow\mathbb{R}$.}\\

\noindent So, we have the corresponding symplectic form
\begin{equation}
\Omega (Z_{1},Z_{2})=\langle{\bf J}_{c}Z_{1},Z_{2}\rangle.
\end{equation}

  \item[b)] {\it The system} (\ref{7}) {\it is equivariant with respect to a
one-parameter symmetry group ${\mathcal G}$ spanned by the generator
${\mathfrak g}$. This group ${\mathcal G}$ is assumed to be either
compact or a subgroup of affine translations. We also suppose that the
  perturbation preserves this symmetry.}\\
  \item[c)] {\it The system}~(\ref{7}) {\it has the family of fixed points
  $\phi p_{0}$, where $p_{0}=0$ and $\phi\in\mathcal{G}$, and
  corresponding (heteroclinic) orbits $Z_{0}(x)$ such that
\begin{equation}
\frac{d}{dx}Z_{0}(x)=f_{0}\left(Z_{0}(x)\right),
\end{equation}
and $\lim_{x\to -\infty}Z_{0}(x)=\phi_{1}p_{0}$ as well as $\lim_{x\to
  +\infty}Z_{0}(x)=\phi_{2}p_{0}$, for some
  $\phi_{1},\phi_{2}\in{\mathcal G}$.}
  \end{description}
\end{description}
{\bf (H3):} {\it $f_{1}(Z,x)=A_{1}Z+f(x)+g(Z,x)$, where $A_{1}$ is a linear
operator, $f(x)=f(x+T)$, $g(Z,x)$ is time-periodic with period $T$ and
also satisfies $g(0,x)=0$, $Dg(0,x)=0$}.
\begin{description}
 \item[(H4)]
 \begin{description}
  \item[a)] {\it For $\epsilon=0$ the spectrum $\sigma[\exp(TA)]=\{1,1,e^{\pm\lambda T}\}$,
$\lambda>0$, where $A={\bf J}_{c}^{-1}D^{2}H_{0}(p_{0})$},\\
  \item[b)]
  \begin{description}
   \item[i.] (Hamiltonian case) {\it For $\epsilon>0$ $\sigma[\exp[T(A+\epsilon
A_{1})]]=\{1,1,e^{T\lambda_{\epsilon}^{\pm}}\}$ },\\
   \item[ii.] (Dissipative case) {\it For $\epsilon>0$ $\sigma[\exp[T(A+\epsilon
A_{1})]]=\{1,\lambda^{d},e^{T\lambda_{\epsilon}^{\pm}}\}$, where
$C_{1}\epsilon\leqslant dist(\lambda^{d},|z|=1)\leqslant C_{2}\epsilon$,
$C_{1}>0$, $C_{2}>0$}.
  \end{description}
 \end{description}
\end{description}

Next, one can define the Poincar\'e map $P^{\epsilon}:\mathbb{R}^{4}\rightarrow\mathbb{R}^{4}$ as
\begin{equation}
P^{\epsilon}(Z)=\pi_{1}\Phi_{T}^{\epsilon}(Z,0),
\end{equation}
where $\pi_{1}:\mathbb{R}^{4}\times S^{1}\rightarrow\mathbb{R}^{4}$
denotes the projection onto the first factor. Equivalently, one can define
\begin{equation}
P^{\epsilon}_{x_{0}}(Z)=\pi_{1}\Phi_{T}^{\epsilon}(Z,x_{0}).
\end{equation}
We rewrite the fixed point equation
$P^{\epsilon}(p_{\epsilon})=p_{\epsilon}$ in the form:
\begin{equation}\label{46}
\mathcal{P}^{\epsilon}(p_{\epsilon})=0,
\end{equation}
where $\mathcal{P}^{\epsilon}(Z)=P^{\epsilon}(Z)-Z$, and the
operator $L=D\mathcal{P}^{0}(0)$ is introduced.

\section{Main results}

\noindent
{\bf Lemma 1.}{\it
Let {\bf (H1)}-{\bf (H4)} hold. For $\epsilon$ small,
  there exists a unique group orbit $\phi p_{\epsilon}$ of fixed points of
  the perturbed Poincar\'e map near the group
  orbit $\phi p_{0}$ such that
\[\min_{\phi_1,\,\,\,\phi_2\in\mathcal{G}}\{\phi_{1}
  p_{\epsilon}-\phi_{2}p_{0}\}=\mathcal{O}(\epsilon).\] Equivalently, there is a family of periodic orbits
  $\phi\gamma_{\epsilon}(x)=(\phi p_{\epsilon},\omega x)$ of the perturbed
  system (\ref{6}) near $\phi\gamma_{0}(x)$ for $\phi\in\mathcal{G}$. }\\
\noindent
{\bf Lemma 2.}{\it
For $\epsilon>0$ sufficiently small, we have
$\sigma\left[DP^{\epsilon}(p_{\epsilon})\right]=\{1,1,e^{T\lambda_{\epsilon}^{\pm}}\}$
for the case {\bf (H4b i)} and
$\sigma\left[DP^{\epsilon}(p_{\epsilon})\right]=\{1,\lambda^{d},e^{T\lambda_{\epsilon}^{\pm}}\}$
for the case {\bf (H4b ii)} respectively.}\\

\noindent {\bf Conjecture 3.} {\it Corresponding to the eigenvalues
${e^{T\lambda_{\epsilon}^{\pm}},1}$ there exist invariant manifolds:
$W^{ss}(\gamma_{\epsilon}(x))$ ( the strong stable manifold),
$W^{u}(\gamma_{\epsilon}(x))$ ( the unstable manifold), and
$W^{c}(\gamma_{\epsilon}(x))$ ( the centre manifold) of $p_{\epsilon}$ for the Poincar\'e map $P^{\epsilon}(Z)$
such that

i. $W^{u}_{loc}(\gamma_{\epsilon}(x))$ and
$W^{ss}_{loc}(\gamma_{\epsilon}(x))$ are tangent to the eigenspaces of
$e^{T\lambda_{\epsilon}^{\pm}}$ respectively at
$\gamma_{\epsilon}$, while $W^{c}_{loc}(\gamma_{\epsilon}(x))$ is at
the same point tangent to the eigenspace corresponding to unity
eigenvalue ( double in the case of Hamiltonian perturbations).
Their global analogues are obtained in the usual way:
\begin{equation}
\begin{array}{l}
W^{ss}(\gamma_{\epsilon}(x))=\bigcup_{x\leq
  0}\Phi^{\epsilon}_{x}W^{ss}_{loc}(\gamma_{\epsilon}(x)),\\
W^{u}(\gamma_{\epsilon}(x))=\bigcup_{x\geq
  0}\Phi^{\epsilon}_{x}W^{u}_{loc}(\gamma_{\epsilon}(x)).
\end{array}
\end{equation}

ii. they are invariant under $P^{\epsilon}(\cdot)$;

iii. $W^{ss}(\gamma_{\epsilon}(x))$ and $W^{u}(\gamma_{\epsilon}(x))$
are $C^{r}$ $\mathcal{O}(\epsilon)$-close to
$W^{s}(\gamma_{0}(x))$ and $W^{u}(\gamma_{0}(x))$ respectively.}

\noindent
{\bf Lemma 4.}{\it
Let $\left(Z^{s,u}_{\epsilon}(x,x_{0}),\omega
  x\right)$ be orbits lying in
$W^{s,u}(\gamma_{\epsilon}(x))$ and originating in an
$\mathcal{O}(\epsilon)$-neighbourhood of
$(Z_{0}(-x_{0}),0)$. Then the following expressions hold with
uniform validity in the indicated time intervals:
\begin{equation}
\begin{array}{l}
Z^{s}_{\epsilon}(x,x_{0})=Z_{0}(x-x_{0})+\epsilon
y^{s}_{\epsilon}(x,x_{0})+\mathcal{O}(\epsilon^{2}),\mbox{
  }x\in[x_{0},\infty),\\
Z^{u}_{\epsilon}(x,x_{0})=Z_{0}(x-x_{0})+\epsilon
y^{u}_{\epsilon}(x,x_{0})+\mathcal{O}(\epsilon^{2}),\mbox{
  }x\in(-\infty,x_{0}],
\end{array}
\end{equation}
where $y^{s,u}_{\epsilon}$ satisfy the first variational equation}:
\begin{equation}
\frac{dy}{dx}={\bf J}_{c}^{-1}D^{2}H_{0}(Z_{0}(x-x_{0}))y+\epsilon
f_{1}(Z_{0}(x-x_{0}),\omega x).
\end{equation}

\noindent We introduce Melnikov function as:
\begin{equation}\label{16}
M(x_{0})=\int_{-\infty}^{\infty}DH_{0}(Z_{0}(x))\cdot
f_{1}(Z_{0}(x),x+x_{0})dx=\int_{-\infty}^{\infty}\Omega\left(f_{0}(Z_{0}(x)),f_{1}(Z_{0}(x),x+x_{0}\right)dx.
\end{equation}

\noindent
{\bf Theorem 1.}{\it
Suppose $M(x_{0})$ has simple zeros. Then for
$\epsilon>0$ sufficiently small $W^{s}(\gamma_{\epsilon})$ and
$W^{u}(\gamma_{\epsilon})$ intersect transversely.}\\

This result implies via the Smale-Birkhoff theorem
\cite{6} the appearance of a horseshoe near
the saddle point of the perturbed Poincar\'e map, what results in a
chaotic dynamics in the corresponding region of the phase space.

\section{Example}

We consider the perturbed generalised Korteweg-de Vries equation
\cite{1,2}:
\begin{equation}\label{17}
U_{t}+\Delta U_{x}+\alpha U^{p}U_{x}+U_{x x x}=\epsilon f_{x}(U,U_{x},U_{t},x),
\end{equation}
where the perturbation is assumed to be periodic of period $T$ in its
last argument: $f(x)=f(x+T)$. It can be rewritten in a multi-symplectic
form as
\begin{equation}
{\bf M}Z_{t}+{\bf K}Z_{x}=\nabla S,\mbox{ }Z=\left(
\begin{array}{c}
U\\
V\\
W\\
\Phi
\end{array}
\right)\in\mathbb{R}^4,
\mbox{ }x\in\mathbb{R}
\end{equation}
with
\begin{equation}
{\bf M}=\left(
\begin{array}{cccc}
0&0&0&-1\\
0&0&0&0\\
0&0&0&0\\
1&0&0&0
\end{array}
\right),\mbox{ }
{\bf K}=\left(
\begin{array}{cccc}
0&-1&0&0\\
1&0&0&0\\
0&0&0&-1\\
0&0&1&0
\end{array}
\right),
\end{equation}
and
\begin{equation}
S=\frac{1}{2}V^2-\frac{1}{2}UW+\frac{1}{2}\Delta U^{2}+\frac{\alpha}{(p+1)(p+2)}U^{p+2}.
\end{equation}
The unperturbed solution is defined as:
\begin{equation}
\left\{
\begin{array}{l}
U_{0}(x)=2b+\frac{3}{\alpha}K^{2}\sech^{2}\left(\frac{Kx}{2}\right),\\
V_{0}(x)=-\frac{3}{\alpha}K^{3}\sinh\left(\frac{Kx}{2}\right)\sech^{3}\left(\frac{Kx}{2}\right),\\
W_{0}(x)=2a+4b(\Delta+\alpha b)+\frac{3}{\alpha}cK^{2}\sech^{2}\left(\frac{Kx}{2}\right),\\
\Phi_{0}(x)=\frac{3}{\alpha}K\left[1+\tanh\left(\frac{Kx}{2}\right)\right],
\end{array}
\right.
\end{equation}
Here $K=\sqrt{c-\Delta-2\alpha b}$. Perturbation can be represented in a
multi-symplectic setting as:
\begin{equation}
\epsilon f_{1}(Z,x)=\left(
\begin{array}{c}
0\\
\epsilon \tilde{f}(Z,x)\\
0\\
0
\end{array}
\right).
\end{equation}
Here $\tilde{f}(Z,x)=f(U,U_{x},-cU_{x},x)$ with $U$ and its derivatives
being components of $Z$. Therefore Melnikov function (\ref{16}) for
the system (\ref{17}) yields
\begin{equation}
M(x_{0})=\int_{-\infty}^{\infty}V_{0}(x)\tilde{f}(Z_{0}(x),x+x_{0})dx.
\end{equation}
For the one-harmonic dissipative driving force this Melnikov function
will have simple zeros \cite{2}, and therefore ona can
conclude the chaotic dynamics of $U$.

\section{Conclusions}

We have considered a modification of Melnikov's method, which can be
used for the analysis of chaotic behaviour of travelling wave
solutions to multi-symplectic PDEs. The results are illustrated with the
example of the perturbed KdV equation.

\subsection*{Acknowledgements}

This work was partially supported by the ORS Scholarship from the
Universities UK. The author is grateful to Gianne Derks for help and
useful comments.

\end{document}